\pdfoutput=1

\documentclass[11pt,twoside,a4paper,cmspaper,final,collab]{cms-tdr}
\def\svnVersion{119714}\def\cmsCernNo{2012-043}\def\cmsCernDate{\today}\def\cmsMessage{Submitted to the European Physical Journal C}

\begin{document}\cmsNoteHeader{HIN-10-005}

\hyphenation{had-ron-i-za-tion}
\hyphenation{cal-or-i-me-ter}
\hyphenation{de-vices}

\RCS$Revision: 119128 $
\RCS$HeadURL: svn+ssh://alverson@svn.cern.ch/reps/tdr2/papers/HIN-10-005/trunk/HIN-10-005.tex $
\RCS$Id: HIN-10-005.tex 119128 2012-05-01 02:14:59Z sungho $
\newcommand{\xt} {\ensuremath{x_{\mathrm{T}}}\xspace}
\newcommand{\microbinv} {\mbox{\ensuremath{\,\mu\text{b}^{-1}}}\xspace}
\newcommand{\nanobinv} {\mbox{\ensuremath{\,\text{n}\text{b}^{-1}}}\xspace}
\newcommand{\RAA} {\ensuremath{R_{\mathrm{AA}}}\xspace}
\newcommand{\TAA} {\ensuremath{T_{\mathrm{AA}}}\xspace}
\newcommand{\RCP} {\ensuremath{R_{\mathrm{CP}}}\xspace}
\newcommand{\sNN} {\ensuremath{\sqrt{s_{_{\mathrm{NN}}}}}\xspace}
\newcommand{\Npart} {\ensuremath{N_{\text{part}}}\xspace}
\newcommand{\Ncoll} {\ensuremath{N_{\text{coll}}}\xspace}
\newcommand{\HYDJET} {\textsc{hydjet}\xspace}
\newcommand{\PYQUEN} {\textsc{pyquen}\xspace}
\newcommand{\AMPT} {\textsc{ampt}\xspace}

\providecommand{\rd}{\ensuremath{\mathrm{d}}} 
\providecommand{\pthat}{\ensuremath{\hat{p}_\mathrm{T}}}
\newlength\cmsFigWidth\setlength\cmsFigWidth{0.48\textwidth}
\newlength\cmsFigAdjWidth
\ifthenelse{\boolean{cms@external}}{\setlength\cmsFigAdjWidth{\cmsFigWidth}}{\setlength\cmsFigAdjWidth{0.65\textwidth}}
\cmsNoteHeader{HIN-10-005} 
\title{Study of high-\pt charged particle suppression in PbPb compared to \Pp\Pp\ collisions
at $\sqrt{s_{_{\mathrm{NN}}}}=2.76\TeV$}

\date{\today}

\abstract{
The transverse momentum spectra of charged particles have been measured in \Pp\Pp\ and PbPb collisions at $\sqrt{s_{_{NN}}}=2.76$\TeV
by the CMS experiment at the LHC.
In the transverse momentum range
$\pt=5$--10\GeVc,
the charged particle yield in the most central PbPb collisions
is suppressed by up to a factor of 7 compared to the \Pp\Pp\ yield scaled by
the number of incoherent nucleon-nucleon collisions.
At higher $\pt$, this suppression is significantly reduced, approaching
roughly a factor of 2 for particles with \pt in the range
$\pt=40$--100\GeVc.
}

\hypersetup{%
pdfauthor={CMS Collaboration},%
pdftitle={Study of high-pT charged particle suppression in PbPb compared to pp collisions at sqrt(sNN)=2.76 TeV},%
pdfsubject={CMS},%
pdfkeywords={CMS, physics, heavy ions}}

\maketitle 

\section{Introduction}

The charged particle spectrum at large transverse momentum (\pt), dominated by hadrons originating
from parton fragmentation, is an important observable for studying the properties
of the hot, dense medium produced in high-energy heavy-ion collisions.
The study of the modifications of the \pt spectrum in PbPb compared to \Pp\Pp\ collisions at the same collision
energy can shed light on the detailed mechanism by which hard partons
lose energy traversing the medium~\cite{dEnterria:2009am,CasalderreySolana:2007zz}, complementing
recent studies of jet quenching and fragmentation properties using fully
reconstructed jets~\cite{jetquenchingATLAS,jetquenchingCMS}.

Using data collected by the Compact Muon Solenoid (CMS) experiment at the LHC,
this paper presents measurements of charged particle yields as a function of \pt
and event centrality in PbPb collisions at a center-of-mass energy per nucleon pair
$\sqrt{s_{NN}} = 2.76\TeV$.
The PbPb charged particle spectra are compared to the corresponding
\pt-differential cross sections measured in \Pp\Pp\ collisions at the same center-of-mass energy,
a measurement that follows closely the analysis described in Ref.~\cite{ppreferenceCMS}.
Charged tracks are measured in the pseudorapidity range $|\eta|<1$, where $\eta=-\ln[\tan(\theta/2)]$,
with $\theta$ the polar angle of the track with respect to the counterclockwise beam direction.

The measurements are motivated by lower-energy results~\cite{Adams:2005dq,Adcox:2004mh,Arsene:2004fa,Back:2004je}
from the Relativistic Heavy Ion Collider (RHIC), where high-\pt particle production was
found to be strongly suppressed relative to expectations from an independent superposition of
nucleon-nucleon collisions. This observation is typically expressed in terms of
the nuclear modification factor,
\begin{equation}
\RAA(\pt) = \frac{\rd^{2}N_{\text{ch}}^{\mathrm{AA}}/\rd\pt\,\rd\eta}{\left<\TAA\right> \rd^{2}\sigma_{\text{ch}}^{\Pp\Pp}/\rd\pt\,\rd\eta},
\label{eqn:def_raa}
\end{equation}
where $N_{\text{ch}}^{\mathrm{AA}}$ and $\sigma_{\text{ch}}^{\Pp\Pp}$ represent the charged
particle yield per event in nucleus-nucleus (AA)
collisions and the charged particle cross section in \Pp\Pp\ collisions, respectively.
In order to compare the yield of high-\pt charged particles produced in PbPb and \Pp\Pp\ collisions,
a scaling factor, the nuclear overlap function $\TAA$, is needed to provide a proper normalization at a given
PbPb centrality. This factor is computed as the ratio between the number of binary nucleon-nucleon
collisions $\Ncoll$, calculated from the Glauber model of the nuclear collision geometry~\cite{glauber},
and the inelastic nucleon-nucleon (NN) cross section $\sigma^{\mathrm{NN}}_{\text{inel}}=64 \pm 5$\,mb
at $\sqrt{s}=2.76$\,\TeV~\cite{PDBook}. It can be interpreted as the NN-equivalent
integrated luminosity per collision at any given PbPb centrality. The mean of the nuclear overlap
function $\left<\TAA\right>$, averaged over a given centrality bin, is used to determine the nuclear
modification factor at that PbPb centrality.

In addition, the centrality dependence of the PbPb spectrum can also
be examined through the \TAA-scaled ratio of spectra in central and peripheral bins,
\begin{equation}
\RCP(\pt) = \frac{(\rd^{2}N_{\text{ch}}^{\mathrm{AA}}/\rd\pt\,\rd\eta)/\left<\TAA\right> \: [\text{central}]}
{(\rd^{2}N_{\text{ch}}^{\mathrm{AA}}/\rd\pt\,\rd\eta)/\left<\TAA\right> \: [\text{peripheral}]}.
\label{eqn:def_rcp}
\end{equation}

In the absence of initial- and/or final-state effects on the PbPb \pt spectrum, the factors \RAA and
\RCP at high \pt are unity by construction. However, as observed first at RHIC in 200\GeV AuAu
collisions~\cite{Adams:2005dq,Adcox:2004mh,Arsene:2004fa,Back:2004je} and later by
ALICE in 2.76\TeV PbPb collisions~\cite{aliceRAA}, the yield of $\pt \sim 5$--10\GeVc charged particles is
suppressed in the most central heavy-ion collisions by up to a factor of five
compared to that in \Pp\Pp\ collisions.
The CMS measurement presented in this paper confirms these results with improved experimental
uncertainties and extends the measured transverse momentum range to 100\GeVc.

\section{Data sample and analysis procedures}
\label{sec:dataAndAnalysis}

This measurement is based on $\sqrt{s_{_{NN}}}=2.76$\TeV PbPb data samples corresponding to integrated
luminosities of 7\microbinv and 150\microbinv, collected by the CMS experiment in 2010 and 2011, respectively.
The \Pp\Pp\ reference measurement uses a data sample collected in $\sqrt{s}=2.76$\TeV collisions
in the 2011 LHC run, corresponding to an integrated luminosity of 230\nanobinv.

A detailed description of the CMS detector can be found in Ref.~\cite{JINST}.
The central feature of the CMS apparatus is a superconducting solenoid
of 6\,m internal diameter, providing an axial magnetic field of 3.8\,T.
Immersed in the magnetic field are the pixel tracker, the silicon strip
tracker, the lead-tungstate crystal electromagnetic calorimeter
(ECAL), and the brass/scintillator hadron calorimeter (HCAL).
Muons are measured in gas ionization detectors embedded in the steel return yoke.
The tracker consists of 1440 silicon pixel and 15\,148 silicon strip
detector modules and measures charged particle trajectories within the nominal
pseudorapidity range $|\eta|< 2.4$. The pixel tracker consists of three
53.3\cm long barrel layers and two endcap disks on each side of the barrel
section. The innermost barrel layer has a radius of 4.4\cm, while for the
second and third layers the radii are 7.3\cm and 10.2\cm, respectively.
The tracker is designed to provide a track impact parameter resolution of about
100\micron and a transverse momentum resolution of about 0.7 (2.0)\% for
1 (100)\GeVc charged particles at normal incidence ($\eta=0$)~\cite{CMSTDR1}.
The beam scintillator counters (BSCs) are located at a distance of 10.86\,m from the nominal
interaction point (IP),
one on each side, and cover the $|\eta|$ range from 3.23 to 4.65.
Each BSC is a set of $16$ scintillator tiles. The BSC elements provide hit and coincidence rates
with a time resolution of 3\,ns and an average minimum ionising particle detection efficiency of 95.7\%.
The two steel/quartz-fibre hadron forward calorimeters (HF), which extend the calorimetric coverage beyond
the barrel and endcap detectors to the $|\eta|$ region between 2.9 and 5.2,
are used for further offline selection of collision events.
For online event selection, CMS uses a two-level trigger system: a hardware level (L1) and a
software-based higher level (HLT).

A sample of minimum bias events from PbPb collisions was collected, based on a trigger requiring
a coincidence between signals in the opposite sides of either the HF or the BSCs.
To ensure a pure sample of inelastic hadronic collision events, additional offline selections were performed.
These include a beam-halo veto, based on the BSC timing,
an offline requirement of at least 3 towers on each HF with an energy
deposit of more than 3\GeV per tower, a reconstructed vertex, based on at least
two pixel tracks with $\pt > 75$\MeVc, and a rejection
of beam-scraping events, based on the compatibility of pixel cluster shapes with the reconstructed primary vertex.
Further details can be found in Ref.~\cite{jetquenchingCMS}.

\begin{figure}[htbp]
        \centering
        \includegraphics[width=\cmsFigWidth]{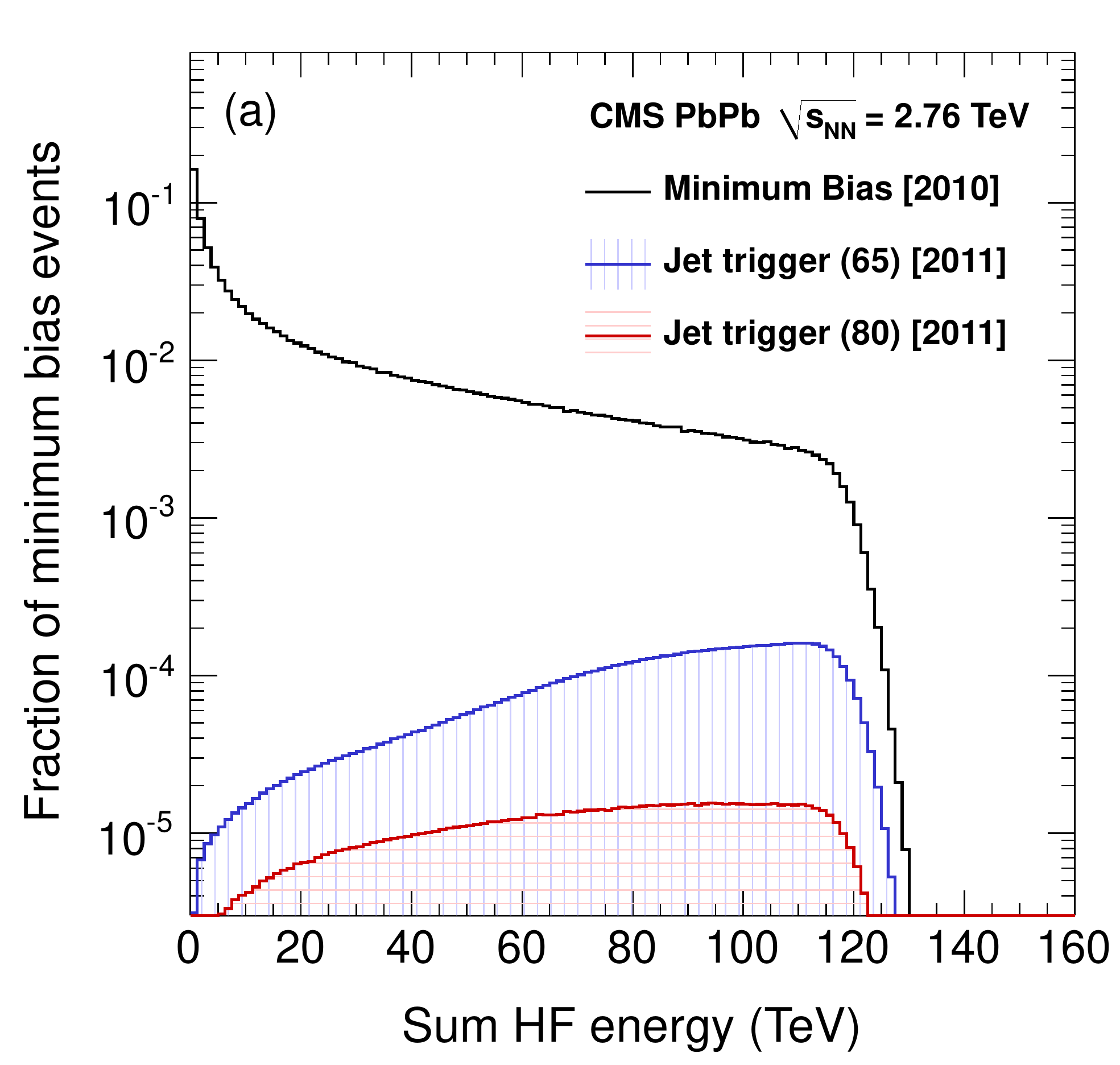}
        \includegraphics[width=\cmsFigWidth]{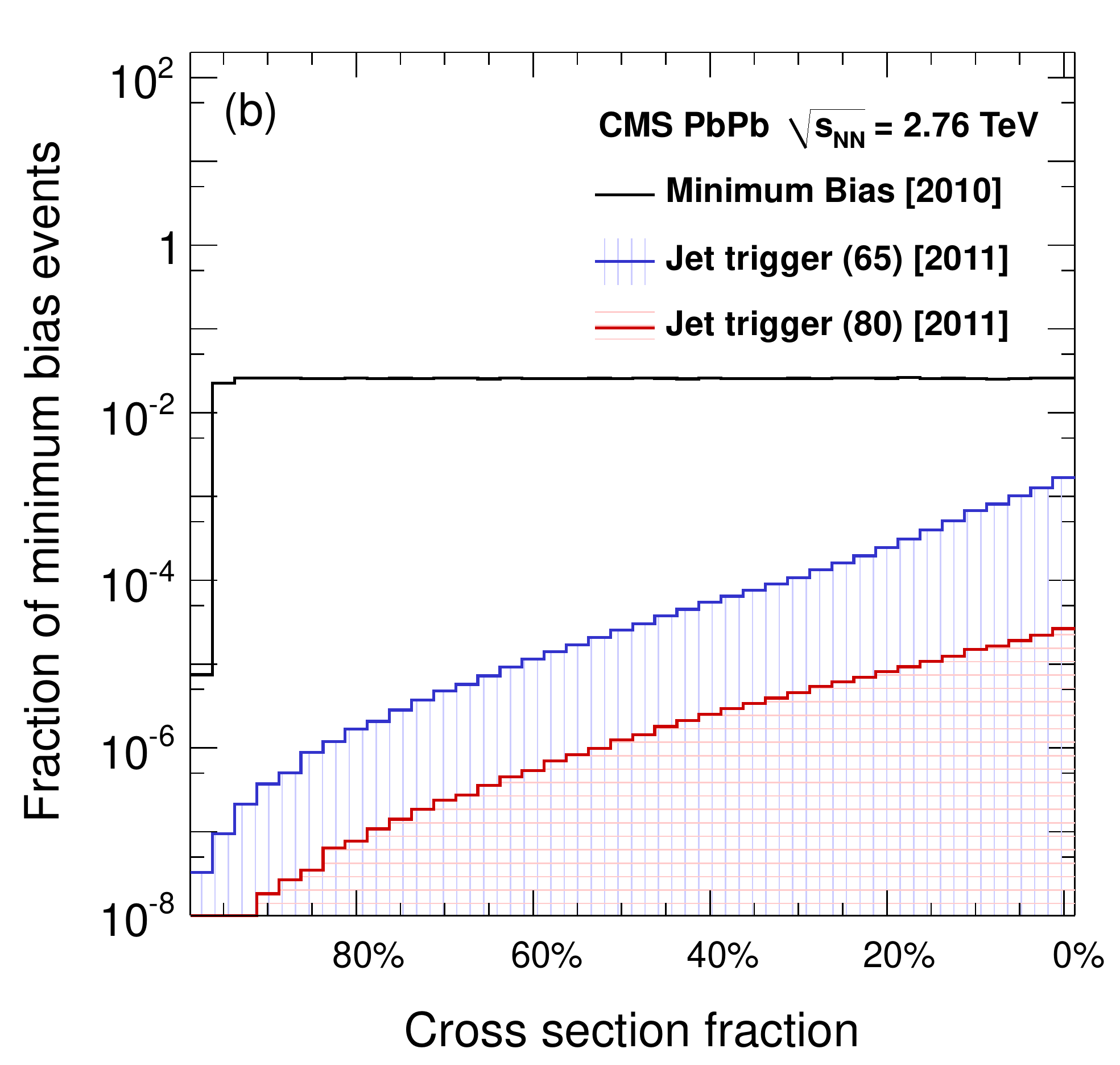}
        \caption{
	(a) Probability distribution of the total HF energy for minimum bias events (black line),
	Jet65-triggered (blue-shaded region), and Jet80-triggered (red-shaded region) events.
        (b) Distribution of the events in bins of fractional cross section for minimum bias (black line),
	Jet65-triggered (blue-shaded region), and Jet80-triggered (red-shaded region) events.
	By convention, 0\% denotes the most central events and 100\% the most peripheral.
	}
    \label{fig:evtCent}
\end{figure}

The collision event centrality is determined from the event-by-event total energy
deposition in both HF calorimeters.
The distribution of this observable in minimum bias events from the 2010 data sample,
shown in Fig.~\ref{fig:evtCent}~(a), is used to divide the event sample into 40 centrality bins,
each corresponding to 2.5\% of the total inelastic cross section.
Figure~\ref{fig:evtCent}~(b) shows the distribution of events according to centrality bin, which is flat by construction for the minimum bias selection,
except in the most peripheral events where the trigger and offline event selection are no longer fully efficient.
Figure~\ref{fig:evtCent} also shows the distributions of the total HF energy
and of the cross-section fraction for the events selected by single-jet triggers
with calibrated transverse energy thresholds of $\ET = 65$\GeV (Jet65) and 80\GeV (Jet80)
from the 2011 data samples.
The reconstruction of calorimeter-based jets in heavy-ion collisions in the online trigger as
well as in the offline analysis is performed with an iterative cone algorithm modified to
subtract the soft underlying event on an event-by-event basis~\cite{Kodolova:2007hd}.
The overall selection efficiency is estimated to be ($97\pm3$)\% based on Monte Carlo (MC)
simulations~\cite{jetquenchingCMS}. For the \Pp\Pp\ analysis, there is an uncertainty
from the estimated number of additional collision interactions in a given beam crossing
(\ie ``event pile-up'') in addition to the uncertainties from the event selection efficiency.
For the PbPb analysis, the uncertainty due to the event pile-up fraction is negligible ($<0.1\%$).

For this analysis, the events are analyzed in six centrality bins: 0--5\% (most central),
5--10\%, \mbox{10--30\%}, 30--50\%, 50--70\%, and 70--90\% (most peripheral).
Details of the centrality determination are described in Ref.~\cite{jetquenchingCMS}.

The event centrality, specified as a fraction of the total inelastic cross section, can be related to
properties of the PbPb collisions such as the
number of nucleons undergoing at least one inelastic collision (\Npart) and the total number of binary nucleon-nucleon collisions (\Ncoll).
The calculation of these properties is based on a Glauber model of the incoming nuclei~\cite{glauber}
and studies of bin-to-bin smearing, caused by finite resolution effects and
evaluated using fully simulated and reconstructed MC events~\cite{jetquenchingCMS}.
The mean and r.m.s.\;of the \Npart, \Ncoll, and \TAA distributions, along with their corresponding systematic
uncertainties, are listed in Table~\ref{tab:centValues}
for the six centrality bins used in this analysis. The uncertainties on the centrality variables are derived from
propagating the uncertainties on the event selection efficiency and on the parameters of the Glauber model.

\begin{table*}[htbp]
\begin{center}
\caption{The average number of participating nucleons (\Npart), number of binary nucleon-nucleon collisions (\Ncoll),
and nuclear overlap function (\TAA) for the centrality bins used in this analysis.
The r.m.s.\;values give the spread over the centrality bins, which are expressed as fractions of the total inelastic PbPb cross section.}
\setlength{\extrarowheight}{3pt}
\label{tab:centValues} \begin{tabular}{|c||c|c||c|c||c|c|}
\hline
Centrality bin & $\left<\Npart\right>$ & r.m.s. & $\left<\Ncoll\right>$ & r.m.s. & $\left<\TAA\right>$ (mb$^{-1}$) & r.m.s. \\
\hline
0--5\%    & 381  $\pm$ 2    & 19.2 & 1660 $\pm$ 130  & 166  & 25.9   $\pm$ 1.06 & 2.60  \\
5--10\%   & 329   $\pm$ 3    & 22.5 & 1310 $\pm$ 110  & 168  & 20.5   $\pm$ 0.94 & 2.62  \\
10--30\% & 224    $\pm$ 4    & 45.9 & 745   $\pm$ 67    & 240  & 11.6   $\pm$ 0.67 & 3.75  \\
30--50\% & 108    $\pm$ 4    & 27.1 & 251   $\pm$ 28    & 101  & 3.92   $\pm$ 0.37 & 1.58  \\
50--70\% & 42.0   $\pm$ 3.5  & 14.4 & 62.8  $\pm$ 9.4   & 33.4 & 0.98   $\pm$ 0.14 & 0.52  \\
70--90\% & 11.4   $\pm$ 1.5  & 5.73 & 10.8  $\pm$ 2.0   & 7.29 & 0.17   $\pm$ 0.03 & 0.11 \\
50--90\% & 26.7   $\pm$ 2.5  & 18.84& 36.9  $\pm$ 5.7   & 35.5 & 0.58   $\pm$ 0.09 & 0.56 \\
\hline
\end{tabular}
\end{center}
\end{table*}

\begin{figure}[htbp]
   \begin{center}
      \includegraphics[width=0.9\cmsFigWidth]{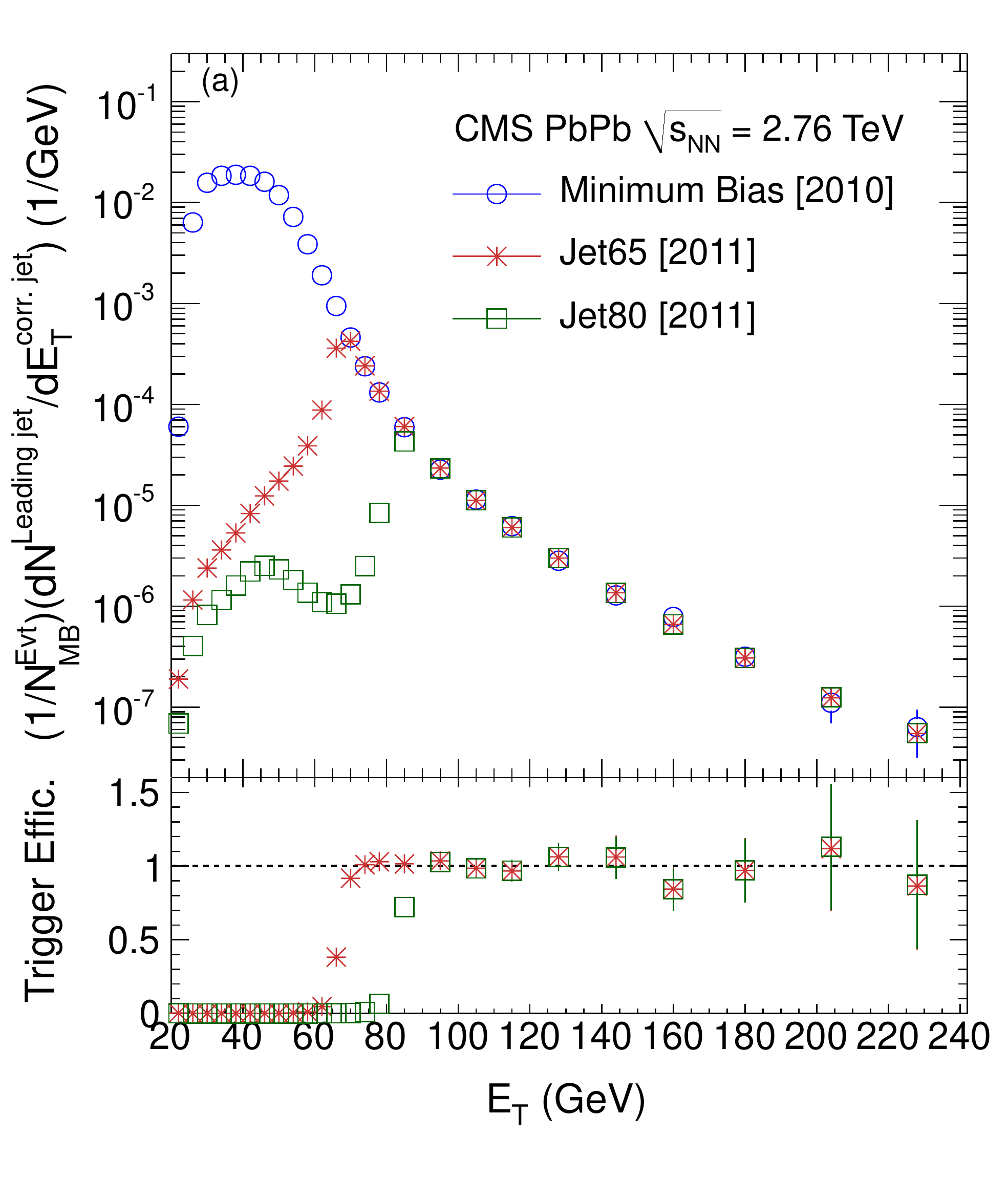}
      \includegraphics[width=0.9\cmsFigWidth]{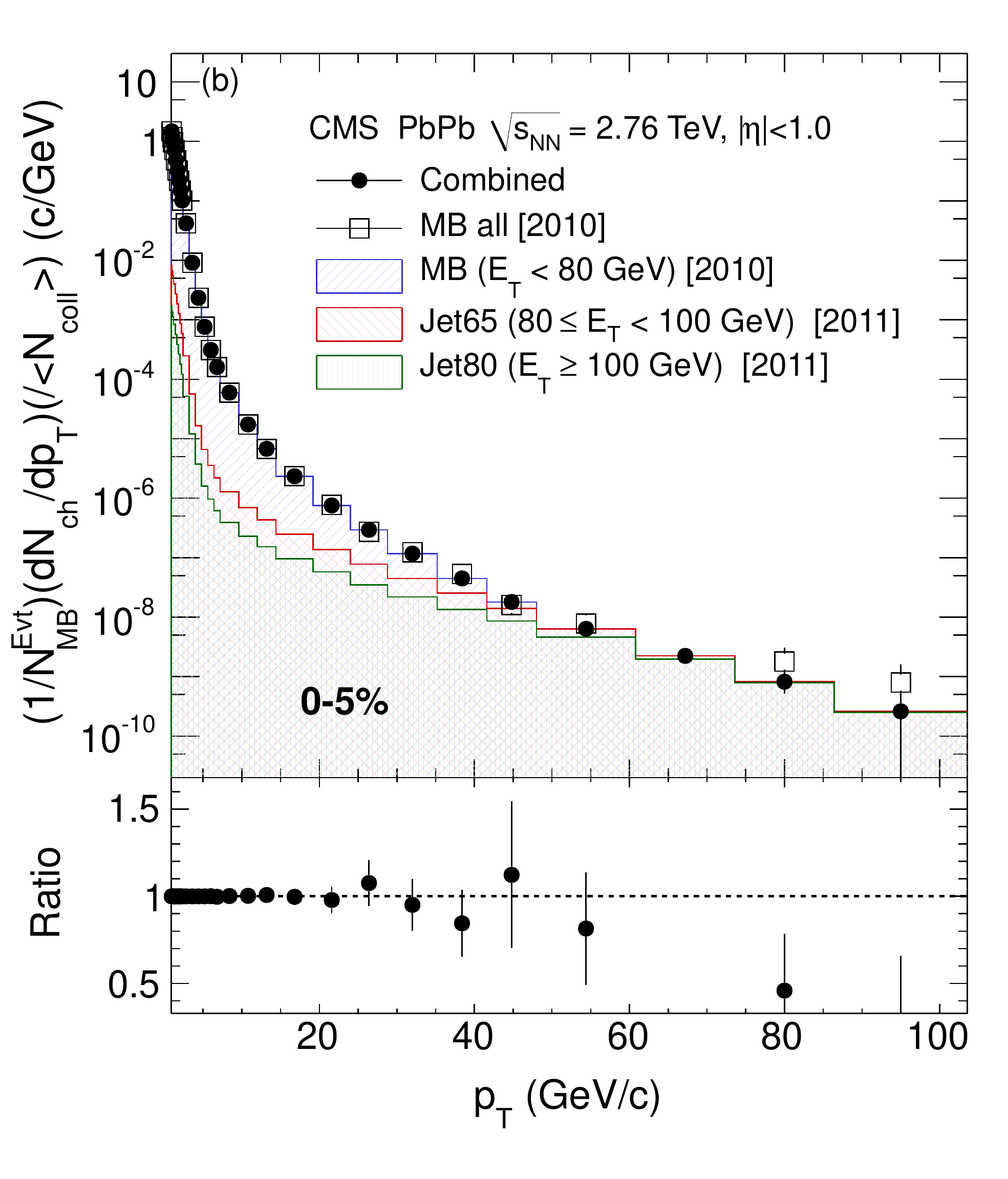}
      \caption{\label{fig:trig}
      (a) Upper panel: Corrected transverse energy \ET of leading jets with $|\eta|<2$ for a
      minimum bias trigger and two jet triggers normalized to the number of
      selected minimum bias events $N^{\text{Evt}}_{\mathrm{MB}}$.
      Lower panel: efficiency curves for the jet triggers with corrected energy thresholds of 65 and 80\GeV.
      (b) Upper panel: The three trigger contributions to the charged particle transverse momentum spectrum and their sum (filled circles)
      for the 0--5\% most central events.  Open squares show the minimum bias spectrum for all values of leading-jet \ET.
      Lower panel: the ratio of the combined spectrum to the minimum bias spectrum.}
   \end{center}
\end{figure}

In order to extend the statistical reach of the \pt spectra in the highly prescaled minimum bias sample,
data recorded in 2011 by unprescaled single-jet triggers, Jet65 and Jet80, are included in the analysis.
The jet \ET thresholds in the trigger are applied
after subtracting the contribution from the underlying event and correcting for the calorimeter response.
Transverse energy distributions of the most energetic reconstructed jet with $|\eta|<2$,
referred to as the leading jet,
are shown in the upper panel of Fig.~\ref{fig:trig}~(a) for the three samples (minimum bias, Jet65, and Jet80)
as a function of corrected \ET, and normalized per minimum bias event.
The distribution for the Jet80 trigger has a peak in the low-\ET region
as a consequence of stricter ECAL and HCAL noise elimination, as well as
a tighter pseudorapidity requirement in the offline leading-jet
selection than in the trigger. This feature is less prominent in the lower threshold Jet65
jet trigger because the rate of noise triggers relative to the rate of true jet triggers
is smaller at lower jet \ET.
The lower panel in Fig.~\ref{fig:trig}~(a) shows the trigger efficiency given by the ratio of each jet-triggered
distribution to that from the immediately looser selection.
The Jet65 (Jet80) trigger becomes fully efficient above $\ET=80$ (100)\GeV.
Following the procedure introduced in the analogous measurement of the charged particle spectra in 0.9 and 7\TeV
pp collisions~\cite{ppreferenceCMS}, the spectra for $|\eta|<1.0$ are calculated separately
in three ranges of leading-jet \ET, below 80\GeV, between 80 and 100\GeV, and above 100\GeV, each corresponding
to a fully efficient trigger path, and then combined to obtain the final result.
Figure~\ref{fig:trig}~(b) shows the contributions from the three ranges to the combined spectrum.
The lower panel of the figure compares the combined spectrum to the minimum
bias spectrum alone, which is in good agreement within statistical uncertainties.
As in the previous analysis~\cite{ppreferenceCMS}, a \pt-dependent normalization uncertainty of \mbox{0--4\%}
is assigned to this procedure of matching the spectra from the different triggered samples.

The reconstruction of charged particles in PbPb collisions, based on hits in the silicon pixel and strip detectors,
is performed similarly to what is done in \Pp\Pp\ collisions~\cite{ppreferenceCMS,TRK-10-001}.
However, some criteria have been fine-tuned to cope with the challenges presented by the
much higher hit density in central PbPb collisions.
First, prior to track reconstruction, the three-dimensional primary
vertex position is fitted from a collection
of pixel-only tracks reconstructed with three hits in the pixel detector and
extrapolating back to a region around the beam spot.
Next, to reduce the random combinatorial background, track candidates
are built from triplet seeds alone, consisting of hits in three layers of the pixel barrel and
endcap detectors.
The seeds from a restricted region within 2\,mm of the primary vertex
are constructed with a minimum \pt of 0.9\GeVc.
Further selections are made on the normalized goodness-of-fit (\ie $\chi^2$) of the track fit
and on the compatibility of the fitted triplet seeds with the primary vertex,
before propagating the seed trajectories through the strip tracker to build fully reconstructed tracks.

To improve the track reconstruction efficiency, two more iterations of the tracking are performed after
removing hits unambiguously belonging to the tracks found in the first iteration. This procedure
is based on the standard \Pp\Pp\ iterative tracking~\cite{TRK-10-001}.
More efficient pp-based pixel-pair and triplet-track seedings
are used in the second and third iterations, respectively.
The tracks found in the later iterations are merged
with the first-iteration tracks after removing any duplicate tracks, based on the fraction of shared hits.
Lastly, the calorimeter (ECAL and HCAL) information is used to improve tracking efficiency at
high \pt (${\gtrsim}30\GeVc$) by requiring looser quality criteria for tracks that
are determined to be calorimeter compatible. This is possible because genuine charged hadron
tracks with high \pt are expected to leave large energy deposits in the calorimeter.
Tracks are matched to the closest
calorimeter cell in ($\eta$, $\phi$), where $\phi$ is the azimuthal angle of the track.
A track is determined to be compatible with the matched calorimeter cell if
the sum of the transverse energy measured by the ECAL and HCAL cells is above a minimum
fraction (30\%) of the track transverse momentum.
Finally, tight quality criteria are imposed for tracks that are incompatible with their matched calorimeter
cell energy. These include requirements of at least 13 hits on the track (counting stereo strip layers separately),
a relative momentum resolution of less than 5\%, a normalized $\chi^2$ of
less than 0.15 times the number of hits, and transverse and longitudinal impact parameters of
less than three times the sum in quadrature of the uncertainties on the impact parameter
and the corresponding vertex position.

Each track is weighted by a factor that accounts for the geometrical acceptance of the detector,
the efficiency of the reconstruction algorithm, the fraction of the tracks for which
a single charged particle is reconstructed as more than one track, the fraction of tracks corresponding
to non-primary charged particles, and the fraction of misidentified tracks that do not
correspond to any charged particle. These correction
factors are applied differentially as functions of pseudorapidity, transverse momentum,
transverse energy of the leading jet, and event centrality.
The various correction terms are estimated based on simulated minimum bias
PbPb events from the \HYDJET~\cite{Lokhtin:2005px} generator.
To improve the statistical precision of the correction factors at high \pt, \HYDJET MC samples are
also mixed at the level of simulated hits with dijet events generated
with different settings of the hard-scattering
scale ($\pthat$=30, 50, 80, 110, and 170\GeVc) from \PYQUEN~\cite{Lokhtin:2005px},
a generator for the simulation of rescattering as well as radiative and collisional
energy loss of hard partons in heavy-ion collisions.

Before applying the tight quality selections on the reconstructed tracks, the charged particle reconstruction efficiency is studied
by inserting simulated pion tracks or \PYQUEN dijet events into two different background samples:
(i) simulated minimum bias \HYDJET events by mixing \GEANTfour~\cite{GEANT4} detector hits,
and (ii) PbPb data events by combining the raw digitized detector signals.
The efficiencies estimated by these two methods agree within 3.0--5.7\% in the range $1 < \pt < 100$\GeVc.
Due to limitations in the data-mixing technique, the two cannot be compared on an equal footing after applying all of the quality cuts,
in particular those involving the consistency of a track with the primary vertex.  However, it is possible to ensure that the distributions on which the selections are made (\ie the $\chi^2$ of the track fit, the distance of
closest approach between track and vertex, the number of hits in the silicon pixel and strip detectors)
are consistent between the data and the MC simulations,
both as a function of \pt and event centrality. To this end, an additional series of checks is performed by varying the
requirements imposed during the
track selection and in the determination of the corresponding MC-based corrections.  The resulting variations in the corrected results
are within the quoted systematic uncertainties.

The fraction of misidentified tracks estimated from simulated events for each leading-jet \ET sample as
a function of track \pt  is checked against an estimate from data that uses the sidebands of the
impact parameter distributions. Studies of simulated events reveal that, at low \pt and
in peripheral events (\eg 50--90\%), the sidebands are dominated by secondaries and products of
weak decays because of their displaced vertex positions. However, in central events (\eg 0--5\%) and at high \pt
they are mostly misidentified tracks.
Based on varying the functional form of the sideband extrapolation under the peak from correctly reconstructed primary tracks,
a 2.5--4.0\% systematic uncertainty is quoted for the fraction of misidentified tracks remaining after all selection cuts.
An additional check is performed for tracks with \pt above 10\GeVc to correlate the reconstructed track momentum with the
energy deposited in the ECAL and HCAL.
The fraction of high-\pt tracks with an atypically small amount of energy deposited in the calorimeters is consistent with the
quoted uncertainty on the misidentification rate.

The tendency for finite bin widths and finite transverse-momentum resolution to deform a steeply falling \pt
spectrum is corrected for in the analysis of the \Pp\Pp\ spectrum~\cite{ppreferenceCMS}.
The higher occupancy in PbPb events than in \Pp\Pp\ events has negligible effect on the momentum resolution.
The resulting 3.0\% systematic uncertainty is dominated by the uncertain shape of the momentum spectrum
at high \pt. For the \RAA and \RCP measurements, a 2.0\% systematic
uncertainty is quoted after subtracting the correlated uncertainty between the PbPb and \Pp\Pp\ \pt spectra,
or between the central and peripheral PbPb \pt spectra.
A summary of all the contributions to the systematic uncertainty affecting the PbPb and \Pp\Pp\ \pt spectra,
and the resulting \RAA and \RCP values, is given in Table~\ref{tab:sys}.

\begin{table*}[htbp]
\caption{Summary of the various contributions to the systematic uncertainties affecting the PbPb and \Pp\Pp\ \pt
spectra, and the nuclear modification factors \RAA and \RCP.}
\centering
\setlength{\extrarowheight}{3pt}
\begin{tabular}{ l c c }
\hline \hline
Source & \multicolumn{2}{c}{Uncertainty [\%]} \\
 & PbPb & \Pp\Pp\ \\
\hline
\ \ \ \ Track reconstruction efficiency & 3.0--5.7  & 2.2--3.6\\                       
\ \ \ \ Non-primary and misidentified tracks & 2.5--4.0  & 1.0--3.2 \\                 
\ \ \ \ Momentum resolution and binning & 3.0 & 0.3--2.7 \\                 
\ \ \ \ Normalization of jet-triggered spectra & 0.0--4.0 & 0.0--6.0 \\   
\ \ \ \ Event selection & 3.0 & 3.5 \\                  
\ \ \ \ Pile-up estimation & $<$0.1 & 1.2 \\
\hline
Total for \pt spectra & 5.8--9.1 & 4.4--9.0 \\
\hline
\ \ \ \ Luminosity & -- & 6.0 \\ 
\hline
\ \ \ \ \TAA determination & 4.1--18.0 & -- \\                                
\hline
Total for \RCP & 6.7--20.0 & -- \\  
\hline
Total for \RAA & 9.9--23.0 & -- \\ 
\hline \hline
\end{tabular}
\label{tab:sys}
\end{table*}

\section{Results}
\label{sec:results}

\begin{figure*}[htbp]
    \centering
    \includegraphics[width=\cmsFigWidth]{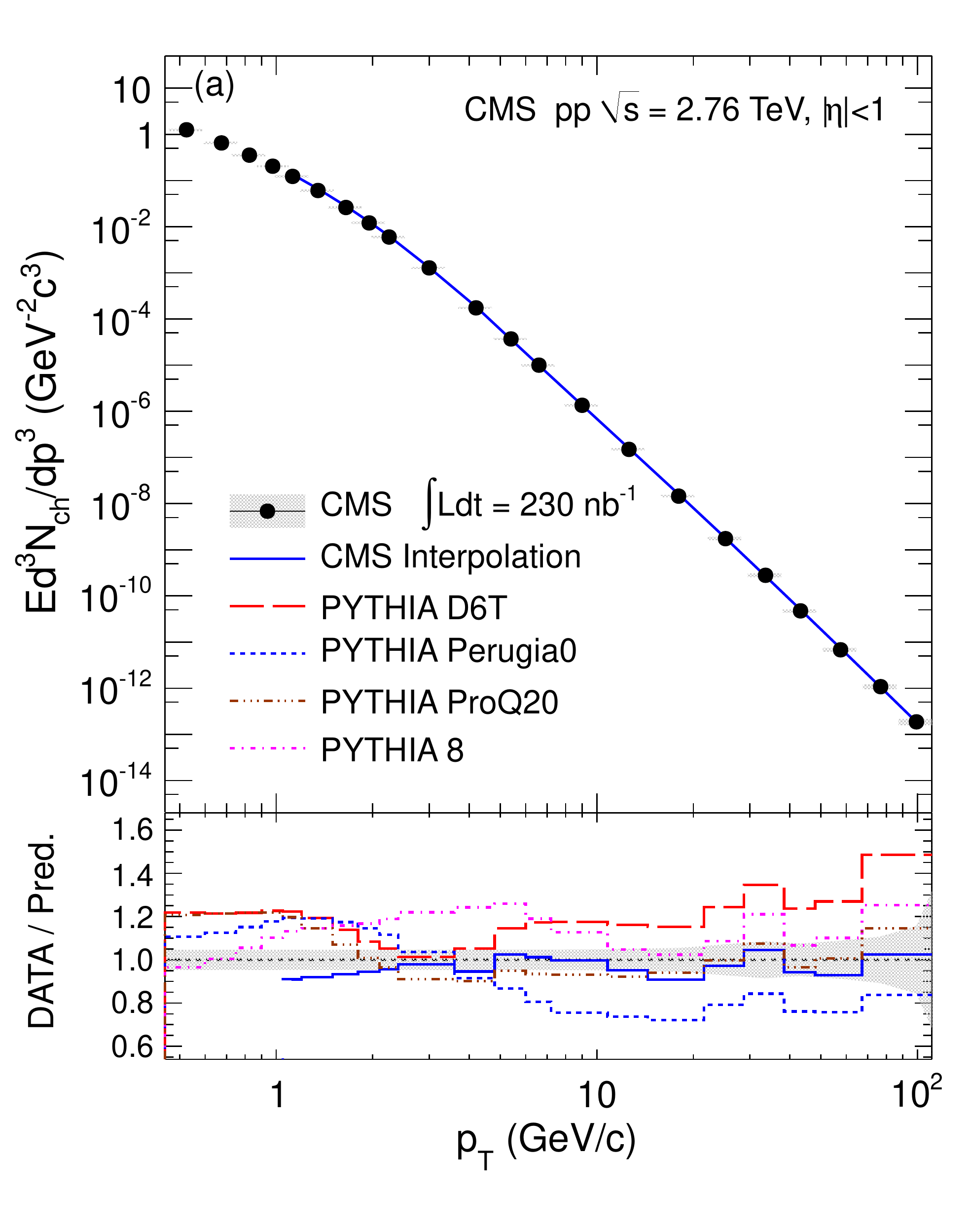}
	\includegraphics[width=\cmsFigWidth]{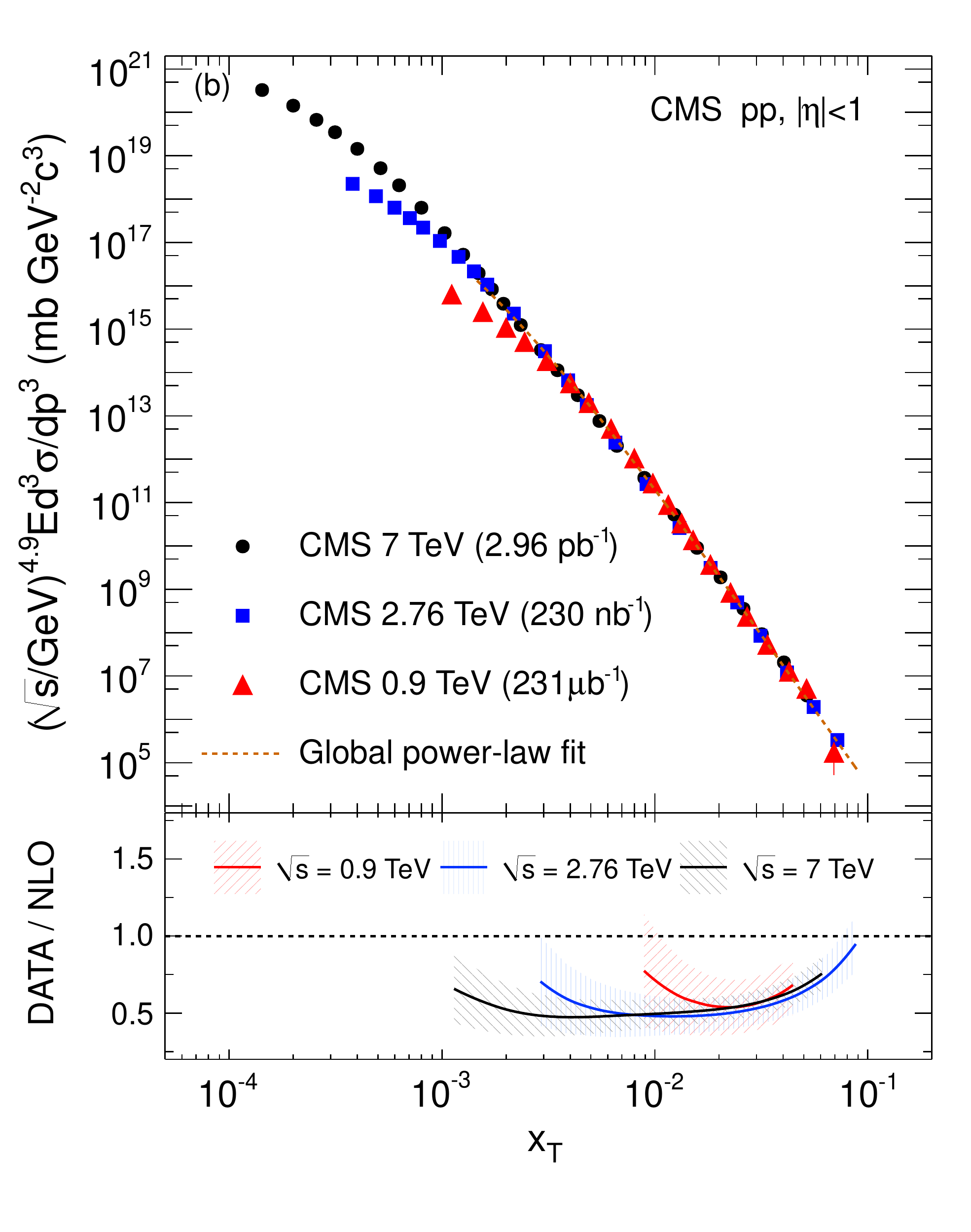}	
      \caption{\label{fig:PP}
      (a) Upper panel: Invariant charged particle differential yield for $|\eta|<1.0$ in \Pp\Pp\ collisions at $\sqrt{s}=2.76$\TeV compared with
      the predictions of four tunes~\cite{Bartalini:2009xx,Sjostrand:2007gs,Skands:2009zm,Buckley:2009bj} of
      the \PYTHIA MC generator and with the CMS interpolated spectrum using data at 0.9 and 7\TeV~\cite{ppreferenceCMS}.
      Lower panel: the ratio of the measured spectrum to the predictions of the four \textsc{pythia} tunes and
      to the interpolated spectrum. The grey band corresponds to the statistical and systematic uncertainties
      of the measurement added in quadrature.
      (b) Upper panel: Inclusive charged particle invariant differential cross sections, scaled by $(\sqrt{s})^{4.9}$, for $|\eta|<1.0$
      as a function of the scaling parameter \xt for CMS data at 0.9 and 7\TeV~\cite{ppreferenceCMS}
      and this analysis at 2.76\TeV. The result is the average of the positive and negative charged particles.
      Lower panel: ratios of the differential cross sections measured at 0.9, 2.76, and 7\TeV
      to those predicted by NLO calculations~\cite{Arleo:2010kw}. The bands show the variations in the
      predictions when changing the factorization scales from 0.5 to 2.0 \pt.
      }
\end{figure*}

The charged particle invariant differential yield ($E\,\rd^{3}N_{\text{ch}} / \rd p^{3} $)
averaged over the pseudorapidity $|\eta|<1.0$
in \Pp\Pp\ collisions is shown in Fig.~\ref{fig:PP}~(a).
The invariant and \pt-differential \Pp\Pp\ cross section is obtained by
normalizing the corresponding yield by the integrated luminosities described in Refs.~\cite{EWK-10-004,EWK-11-001}.
Also shown in Fig.~\ref{fig:PP}~(a) are various generator-level predictions
from the \PYTHIA MC~\cite{Sjostrand:2006za} for different tunes~\cite{Bartalini:2009xx,Sjostrand:2007gs,Skands:2009zm,Buckley:2009bj}, and the ratios of the data to the various MC predictions.
The \Pp\Pp\ measurement is also compared to the empirical global power-law scaling
prediction~\cite{Arleo:2010kw} with an exponent $n=4.9$ determined from the previous CMS measurements~\cite{ppreferenceCMS} by plotting
\ifthenelse{\boolean{cms@external}}{$(\sqrt{s})^{n=4.9} \times E \,\rd^{3}\sigma / \rd p^{3} $}{$(\sqrt{s})^{n=4.9} \, E \,\rd^{3}\sigma / \rd p^{3} $} versus the scaling variable \xt = 2\pt/$\sqrt{s}$,
as shown in Fig.~\ref{fig:PP}~(b).

The \Pp\Pp\ measurement at $\sqrt{s}=2.76$\TeV is consistent with the global power-law fit established
in Ref.~\cite{ppreferenceCMS}. The next-to-leading-order (NLO) prediction~\cite{Arleo:2010kw} for $\sqrt{s}=2.75$ TeV
overestimates the measured cross section by almost a factor of two, as shown in Fig.~\ref{fig:PP}~(b).

\begin{figure}[htb]
   \begin{center}
       \includegraphics[width=\cmsFigAdjWidth]{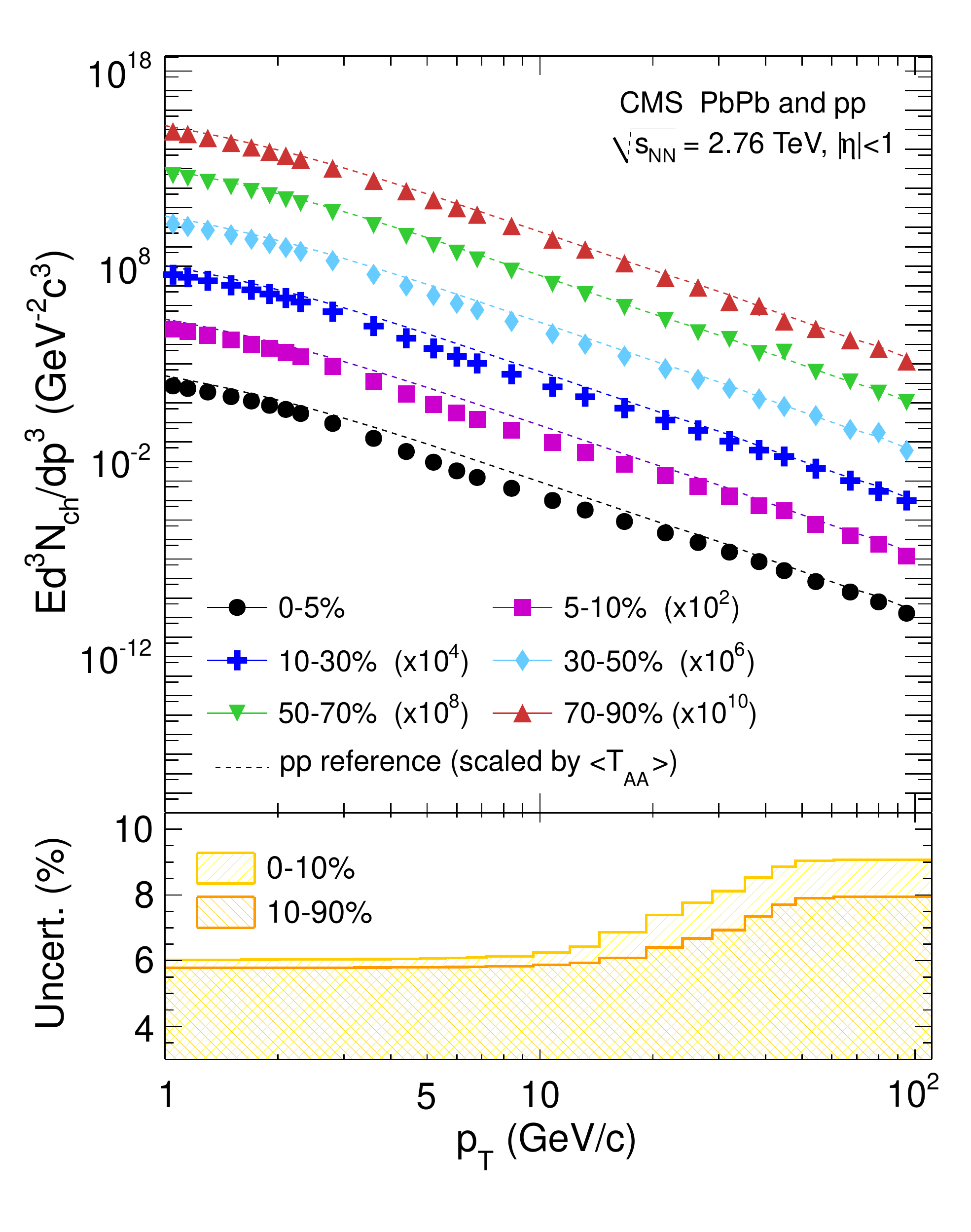}
        \caption{
      Upper panel: Invariant charged particle differential yield in PbPb collisions at 2.76\TeV in bins of
      collision centrality (symbols),
      compared to that of \Pp\Pp\ at 2.76\TeV, normalized by the corresponding \Pp\Pp\ invariant cross sections scaled by the
      nuclear overlap function (dashed lines).
      The spectra for different centrality bins have been scaled by the arbitrary factors shown in the figure,
       for easier viewing.
      The statistical uncertainty is smaller than the marker size for most of the points.
      Lower panel: The average relative systematic uncertainties of the PbPb differential yields
      for the 0--10\% and 10--90\% centrality intervals, as a function of \pt.
	}
     \label{fig:PbPb}
   \end{center}
\end{figure}

The PbPb spectrum is shown for six centrality bins and compared to the measured \Pp\Pp\ reference
spectrum, scaled by the nuclear overlap function, in Fig.~\ref{fig:PbPb}.
For easier viewing, several sets of points have been scaled by the arbitrary factors given in the figure.
By comparing the PbPb measurements to the dashed lines representing the scaled \Pp\Pp\
reference spectrum, it is clear that the charged particle spectrum is strongly suppressed in central
PbPb events compared to \Pp\Pp, with the most pronounced suppression at around 5--10\GeVc.

\begin{figure*}[htb]
   \begin{center}
        \includegraphics[width=0.98\textwidth]{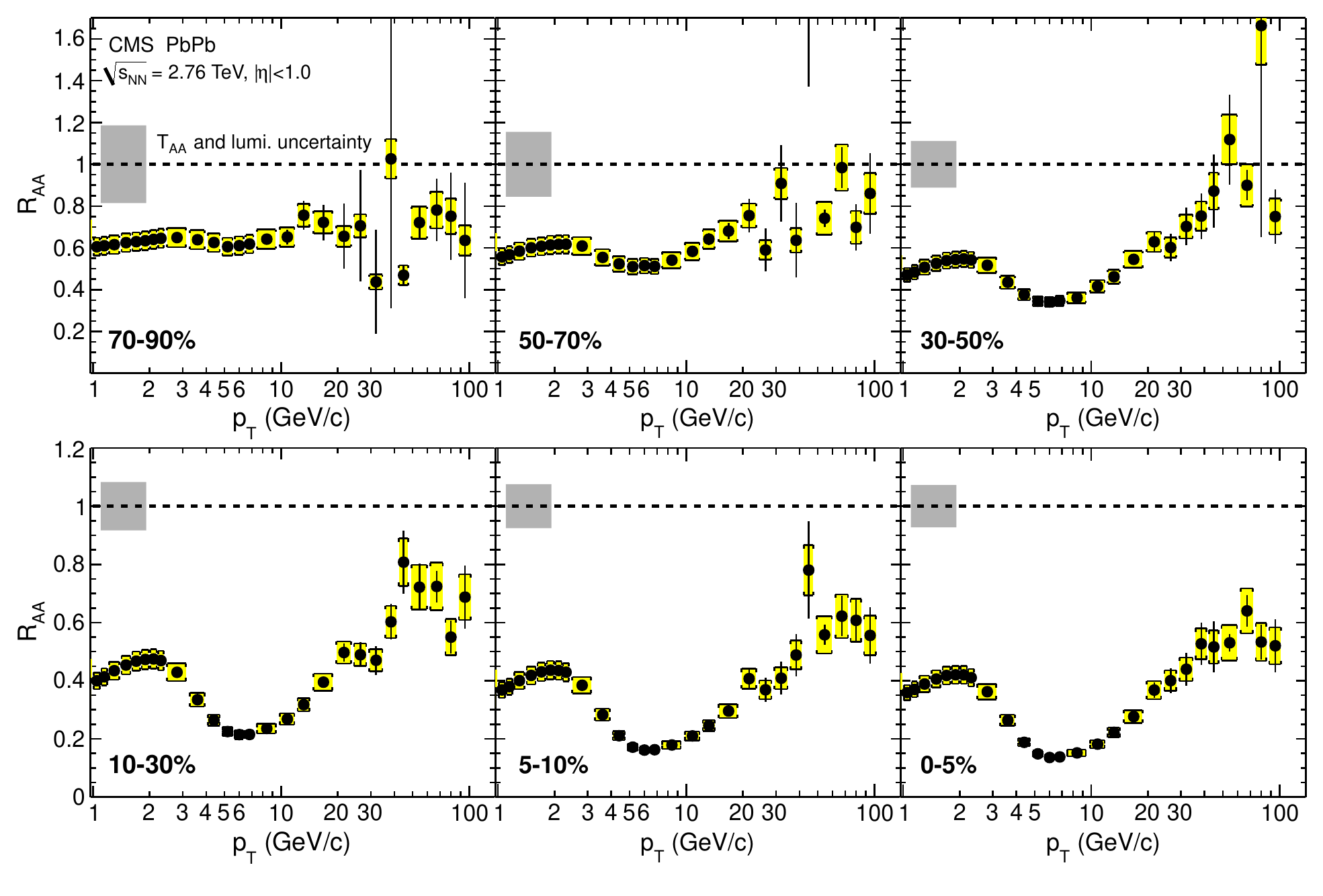}
	\caption{Nuclear modification factor \RAA (filled circles) as a function of \pt for six PbPb centralities.
	The error bars represent the statistical uncertainties and the yellow boxes represent the \pt-dependent systematic uncertainties.
	An additional systematic uncertainty from the normalization of \TAA and the \Pp\Pp\ integrated luminosity, common to all points,
	is shown as the shaded band around unity in each plot.
	}
     \label{fig:specRaaLogX}
   \end{center}
\end{figure*}

The nuclear modification factor \RAA is constructed according to Eq.~(\ref{eqn:def_raa}) by
dividing the PbPb \pt spectrum for each centrality range by the scaled \Pp\Pp\ reference spectrum
(\ie the filled points by the dashed lines in Fig.~\ref{fig:PbPb}). It is presented as a function
of \pt in Fig.~\ref{fig:specRaaLogX} for each of the six centrality bins.
The yellow boxes around the points show the systematic uncertainties, including those from the \Pp\Pp\
reference spectrum, listed in Table~\ref{tab:sys}.  An additional systematic uncertainty from the \TAA normalization,
common to all points and also listed in Table~\ref{tab:sys}, is displayed as the shaded band around unity in each plot.
The statistical uncertainties do not increase monotonically as a function of \pt, as seen most
prominently in the peripheral bins, as a consequence of combining the highly prescaled minimum bias sample
with the two unprescaled jet triggers, as discussed in Section~\ref{sec:dataAndAnalysis}.
In the most peripheral events (70--90\%), a moderate suppression of about a factor of 2 ($\RAA \approx 0.6$)
is observed at low \pt,
with \RAA rising slightly with increasing transverse momentum.
The suppression becomes more pronounced in the more central collisions,
as expected from the increasingly dense final-state system and longer average path-lengths
traversed by hard-scattered partons before fragmenting into final hadrons.
In the 0--5\% centrality bin, \RAA reaches a minimum value of about 0.13 at \pt = 6--7\GeVc.
At higher \pt, the value of \RAA rises and levels off above 40\GeVc at approximately 0.5.
A rising \RAA may simply reflect the flattening of the unquenched nucleon-nucleon spectrum at high \pt
if one assumes a constant fractional energy loss, although the magnitude of the rise varies among
the different theoretical models.

\begin{figure*}[htb]
   \begin{center}
        \includegraphics[width=0.8\textwidth]{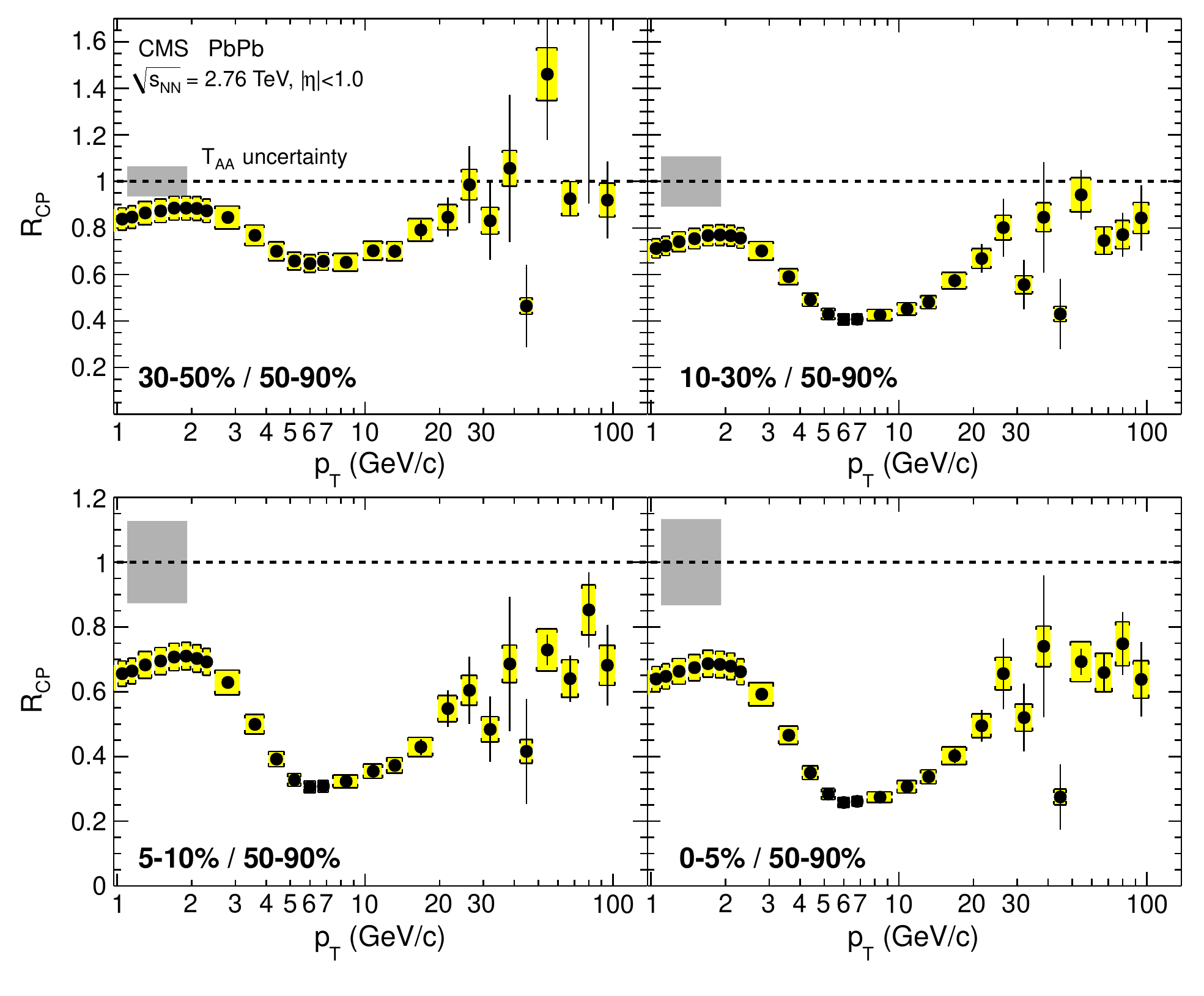}
        \caption{\TAA-scaled ratio of \pt spectra in central and peripheral bins, \RCP, as a function of \pt
	for four PbPb centralities.
        The error bars represent the statistical uncertainties and the yellow boxes the \pt-dependent
	systematic uncertainties.
        An additional systematic uncertainty from the normalization of \TAA, common to all points,
	is shown as the shaded band around unity in each plot.}
     \label{fig:specRcpLogX}
   \end{center}
\end{figure*}

The \TAA-scaled ratio of spectra in central and peripheral bins, \RCP, is constructed according
to Eq.~(\ref{eqn:def_rcp}). The peripheral interval used for the normalization is chosen as the
combined 50--90\% centrality bin to improve the statistical precision at high \pt.  This approach removes
the 4.4--9.0\% systematic uncertainty from the \Pp\Pp\ reference.
Also part of the \TAA uncertainties is
correlated between centrality bins and cancels out in the \RCP ratio.  The resulting values of \RCP for the
four most central bins are shown in Fig.~\ref{fig:specRcpLogX}. The statistical uncertainty of \RCP
does not increase monotonically with \pt for the same reasons as mentioned for \RAA. As in the measurement of \RAA, the \RCP
results show that the \pt spectra in central PbPb collisions are significantly suppressed compared to peripheral
collisions.

\begin{figure*}[htb]
   \begin{center}
      \includegraphics[width=0.8\textwidth]{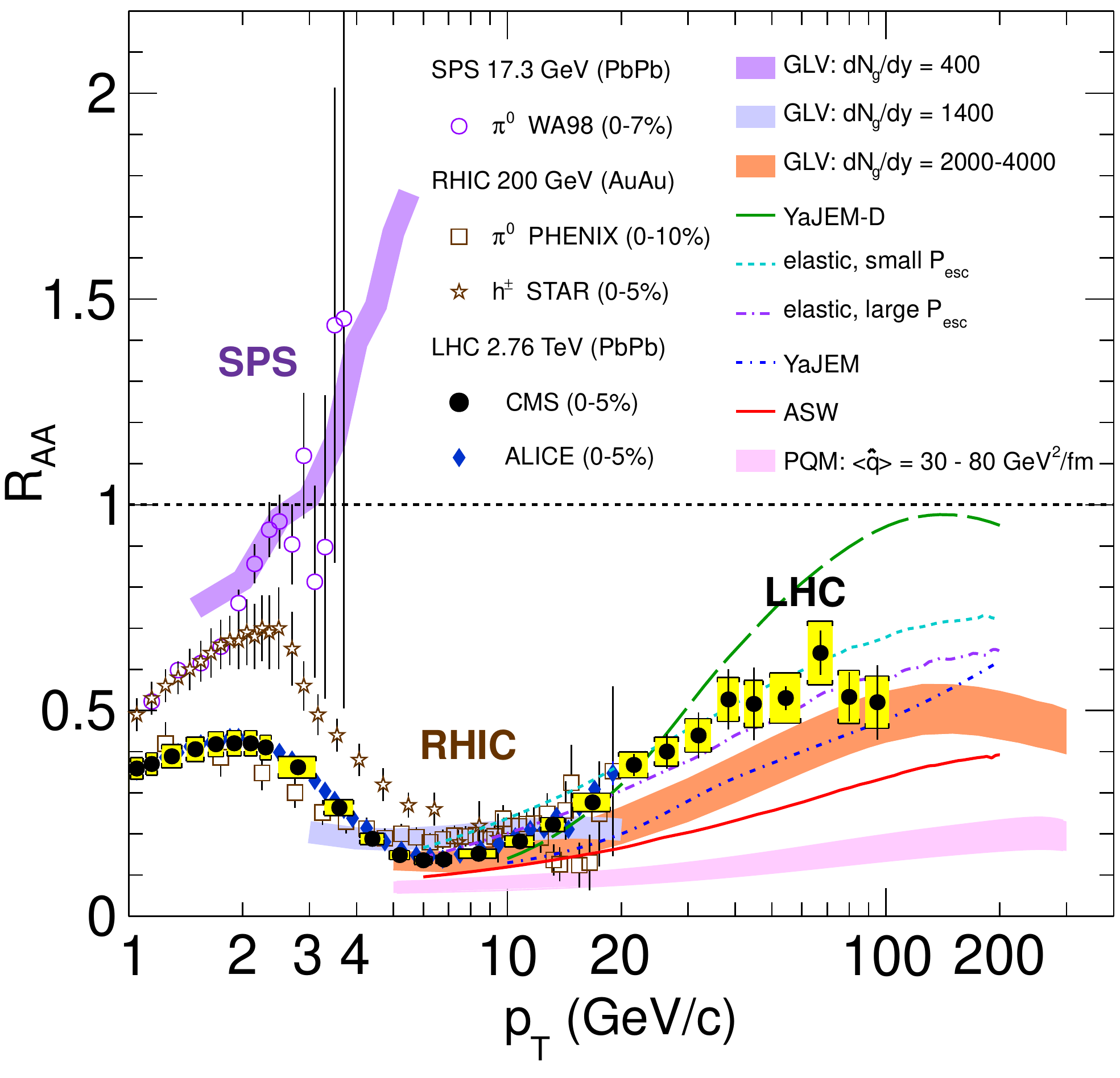}
      \caption{Measurements of the nuclear modification factor \RAA in central heavy-ion collisions at
      three different center-of-mass energies, as a function of \pt, for neutral pions ($\Pgpz$),
      charged hadrons ($h^{\pm}$), and charged particles~\cite{aliceRAA,wa98,wa98revisited,phenix,star},
      compared to several theoretical predictions~\cite{pqm,glv,glv2,Salgado:2003gb,Armesto:2005iq,renk} (see text).
      The error bars on the points are the statistical uncertainties, and the yellow boxes around the CMS points
      are the systematic uncertainties.
      Additional absolute $\TAA$ uncertainties of order $\pm$5\% are not plotted.
      The bands for several of the theoretical calculations represent their uncertainties.}
            \label{fig:raaCMSvsTheoryLinY}
   \end{center}
\end{figure*}

The evolution of the nuclear modification factor with center-of-mass energy,
from the SPS~\cite{wa98,wa98revisited} to RHIC~\cite{phenix,star} and then
to the LHC~\cite{aliceRAA}, is presented in Fig.~\ref{fig:raaCMSvsTheoryLinY}.
Note that RHIC results are shown for both neutral pions and charged hadrons, the latter being less
suppressed below $\pt\approx 8\GeVc$~\cite{phenix,star} possibly due to parton recombination processes that
enhance proton production and thus the overall yield of charged hadrons~\cite{starPID}.
Below $\pt\approx 10\GeVc$, charged hadron production at the LHC is found to be about 50\% more suppressed
than at RHIC, and has a similar suppression value as for neutral pions measured by PHENIX~\cite{phenix}.

The CMS measurement of \RAA presented in this paper for the 0--5\% centrality interval is compared to
the ALICE result~\cite{aliceRAA} in Fig.~\ref{fig:raaCMSvsTheoryLinY}.
Note that the \Pp\Pp\ spectrum measured by CMS at $\sqrt{s}=2.76$\TeV is roughly 5--15\%
higher than the ALICE spectrum obtained by interpolating their 0.9 and 7\TeV spectra~\cite{aliceRAA}.
The two \RAA results are in agreement
within their respective statistical and systematic uncertainties.

The high-\pt measurement of \RAA from this analysis, up to $\pt = 100$\GeVc, is also compared
to a number of theoretical predictions, for the LHC design energy of $\sNN=5.5$\TeV
(PQM~\cite{pqm} with medium transport-coefficient $\langle\hat{q}\rangle =$ 30--80 $\GeV^{2}$/fm
and GLV~\cite{glv,glv2} for various values of the medium gluon pseudorapidity density $\rd N_{g}/\rd y$)
and for the actual collision energy of $\sNN=2.76$\TeV
(ASW~\cite{Salgado:2003gb,Armesto:2005iq} and YaJEM~\cite{renk} including a model for elastic energy loss
parameterized with the $P_{\text{esc}}$ variable).
While most models predict the generally
rising behavior of \RAA that is observed in the data at high \pt, the magnitude of the predicted
slope varies greatly between models, depending on the assumptions for the jet-quenching mechanism.
The new CMS measurement presented here should help in constraining the quenching parameters used in these models
and improve the understanding of parton energy loss in a hot and dense medium.

\section{Summary}

Measurements of the charged particle transverse momentum spectra have been
presented for $\sNN=2.76$\TeV \Pp\Pp\ and PbPb collisions.
The results for the PbPb collisions have been compared to the measured \Pp\Pp\ \pt spectrum
scaled by the corresponding number of incoherent nucleon-nucleon collisions.
The high-\pt yields in central PbPb collisions are significantly suppressed when compared to
peripheral PbPb and \Pp\Pp\ collisions. In the range \pt = 5--10\GeVc, the suppression is stronger
than that seen at RHIC. Beyond 10\GeVc, both \RAA and \RCP show a rising trend, as already
suggested by data from the ALICE experiment, limited to $\pt = 20$\GeVc.
The CMS measurement, with improved statistical precision, clearly shows
that this rise continues at higher \pt, approaching a suppression factor
$\RAA \approx 0.5$--0.6 in the range 40--100\GeVc.
The overall \pt dependence of the suppression can be described by a number of
phenomenological predictions. The detailed evolution of the \RAA rise
from 6 to 100\GeVc depends on the details of the models.
Together with measurements of high-\pt charged hadron azimuthal anisotropies,
inclusive jet spectra, fragmentation functions, and dijet transverse energy balance,
this measurement of the nuclear modification factors as a function of \pt and collision
centrality should help elucidate the mechanism of jet quenching and the properties of
the medium produced in heavy-ion collisions at collider energies.

\section*{Acknowledgements}

\label{sect:acknowledge}

\hyphenation{Bundes-ministerium Forschungs-gemeinschaft Forschungs-zentren} We wish to congratulate our colleagues in the CERN accelerator departments for the excellent performance of the LHC machine. We thank the technical and administrative staff at CERN and other CMS institutes. This work was supported by the Austrian Federal Ministry of Science and Research; the Belgium Fonds de la Recherche Scientifique, and Fonds voor Wetenschappelijk Onderzoek; the Brazilian Funding Agencies (CNPq, CAPES, FAPERJ, and FAPESP); the Bulgarian Ministry of Education and Science; CERN; the Chinese Academy of Sciences, Ministry of Science and Technology, and National Natural Science Foundation of China; the Colombian Funding Agency (COLCIENCIAS); the Croatian Ministry of Science, Education and Sport; the Research Promotion Foundation, Cyprus; the Ministry of Education and Research, Recurrent financing contract SF0690030s09 and European Regional Development Fund, Estonia; the Academy of Finland, Finnish Ministry of Education and Culture, and Helsinki Institute of Physics; the Institut National de Phys\-ique Nucl\'eaire et de Physique des Particules~/~CNRS, and Commissariat \`a l'\'Energie Atomique et aux \'Energies Alternatives~/~CEA, France; the Bundesministerium f\"ur Bildung und Forschung, Deutsche Forschungsgemeinschaft, and Helmholtz-Gemeinschaft Deutscher Forschungszentren, Germany; the General Secretariat for Research and Technology, Greece; the National Scientific Research Foundation, and National Office for Research and Technology, Hungary; the Department of Atomic Energy and the Department of Science and Technology, India; the Institute for Studies in Theoretical Physics and Mathematics, Iran; the Science Foundation, Ireland; the Istituto Nazionale di Fisica Nucleare, Italy; the Korean Ministry of Education, Science and Technology and the World Class University program of NRF, Korea; the Lithuanian Academy of Sciences; the Mexican Funding Agencies (CINVESTAV, CONACYT, SEP, and UASLP-FAI); the Ministry of Science and Innovation, New Zealand; the Pakistan Atomic Energy Commission; the Ministry of Science and Higher Education and the National Science Centre, Poland; the Funda\c{c}\~ao para a Ci\^encia e a Tecnologia, Portugal; JINR (Armenia, Belarus, Georgia, Ukraine, Uzbekistan); the Ministry of Education and Science of the Russian Federation, the Federal Agency of Atomic Energy of the Russian Federation, Russian Academy of Sciences, and the Russian Foundation for Basic Research; the Ministry of Science and Technological Development of Serbia; the Ministerio de Ciencia e Innovaci\'on, and Programa Consolider-Ingenio 2010, Spain; the Swiss Funding Agencies (ETH Board, ETH Zurich, PSI, SNF, UniZH, Canton Zurich, and SER); the National Science Council, Taipei; the Scientific and Technical Research Council of Turkey, and Turkish Atomic Energy Authority; the Science and Technology Facilities Council, UK; the US Department of Energy, and the US National Science Foundation.

Individuals have received support from the Marie-Curie programme and the European Research Council (European Union); the Leventis Foundation; the A. P. Sloan Foundation; the Alexander von Humboldt Foundation; the Belgian Federal Science Policy Office; the Fonds pour la Formation \`a la Recherche dans l'Industrie et dans l'Agri\-culture (FRIA-Belgium); the Agentschap voor Innovatie door Wetenschap en Technologie (IWT-Belgium); the Council of Science and Industrial Research, India; and the HOMING PLUS programme of Foundation for Polish Science, cofinanced from European Union, Regional Development Fund.

\bibliography{auto_generated}   

\cleardoublepage \appendix\section{The CMS Collaboration \label{app:collab}}\begin{sloppypar}\hyphenpenalty=5000\widowpenalty=500\clubpenalty=5000\textbf{Yerevan Physics Institute,  Yerevan,  Armenia}\\*[0pt]
S.~Chatrchyan, V.~Khachatryan, A.M.~Sirunyan, A.~Tumasyan
\vskip\cmsinstskip
\textbf{Institut f\"{u}r Hochenergiephysik der OeAW,  Wien,  Austria}\\*[0pt]
W.~Adam, T.~Bergauer, M.~Dragicevic, J.~Er\"{o}, C.~Fabjan, M.~Friedl, R.~Fr\"{u}hwirth, V.M.~Ghete, J.~Hammer\cmsAuthorMark{1}, M.~Hoch, N.~H\"{o}rmann, J.~Hrubec, M.~Jeitler, W.~Kiesenhofer, M.~Krammer, D.~Liko, I.~Mikulec, M.~Pernicka$^{\textrm{\dag}}$, B.~Rahbaran, C.~Rohringer, H.~Rohringer, R.~Sch\"{o}fbeck, J.~Strauss, A.~Taurok, F.~Teischinger, P.~Wagner, W.~Waltenberger, G.~Walzel, E.~Widl, C.-E.~Wulz
% ...
\vskip\cmsinstskip
\dag:~Deceased\\
1:~~Also at CERN, European Organization for Nuclear Research, Geneva, Switzerland\\
% ...

\end{sloppypar}
\end{document}